\documentclass[prd,amsmath,amssymb]{revtex4}

\usepackage{graphicx}
\usepackage{dcolumn}
\usepackage{bm}

\begin{document}


\vspace*{1.5cm}

\title{ Precision Measurements of the Top Quark Mass at the Tevatron}

 \author{Daniel Whiteson\\ ( On behalf of the CDF and D\O\ collaborations )}
\affiliation{University of Pennsylvania, Philadelphia PA.}

\date{\today}

\begin{abstract}
\vspace*{3.0cm}

We report precision measurements of the top quark mass using events collected by the D\O\ and CDF II detectors from $p\overline{p}$ collisions at $\sqrt s = 1.96$ TeV at the Fermilab Tevatron.    Measurements are presented in multiple decay channels.  In addition, we present a combination of the most precise measurements in each channel to date:
\[ M_{top} = 172.5 \pm 1.3_{stat} \pm 1.9_{syst}\ {\textrm GeV}/c^2  \]

\end{abstract}


\maketitle

\section{\label{sec:Intro}Introduction}
Precision measurements of the top quark mass, $M_t$, combined with precision 
measurement of the $W$ boson mass, place constraints on the masses of particles to which the top quark contributes radiative corrections, including the unobserved Higgs boson~\cite{higgs} and particles in extensions to the standard model~\cite{susy}. At the Tevatron, top quarks are primarily produced in pairs, each decaying almost immediately to a $W$ and a $b$ . Top quark decays are classified according to the decay of the $W$ boson: hadronic ($t\overline{t} \to \overline{b}qq'bqq'$),  lepton+jets ($t\overline{t} \to \overline{b}\ell^{-}\overline{\nu}_{\ell}bqq'$) and dilepton ($t\overline{t} \to \overline{b}\ell^{-}\overline{\nu}_{\ell}b\ell'^{+}\nu'_\ell$).

Measurements in run I~\cite{a,b,c,d} used $\int \mathcal{L} dt \approx 100$ pb$^{-1}$ of data and were dominated by statistical uncertainties due to the small collected samples. In the larger integrated luminosities of run II, measurements in the lepton+jets channels are dominated by systematic uncertainties. Measurements in the dilepton and hadronic channels are approaching similar precision, allowing for a comparison of measurements between channels.

\section{ Sample and Mass Measurement Techniques}

Top quark candidate events in the lepton+jet sample are distinguished by the presence of a high $p_{T}$ electron or muon, four or more jets, significant missing transverse energy from the escaping neutrino; requiring the presence of at least one tagged $b$-jet greatly reduces the backgrounds from $W$+jets production and multi-jet processes.   

Dilepton candidates have two high $p_{T}$ leptons, at least two jets and significant missing energy.  The most significant backgrounds are $Z$+jets decays, $W$+jets where a jet is misidentified as a lepton and diboson production in association with jets.

Techniques to extract the top quark mass from the candidate events translate the kinematic information in each event to a top mass likelihood.   Template methods parameterize the likelihood and extract the values of the parameters from large samples of simulated events at varying top masses. Parameterizations are typically in terms of a reconstructed mass per event.  Matrix-element approaches approximate a direct calculation of the likelihood by convoluting the matrix-element with detector resolution functions.

\section{Lepton+Jets Measurements}

\begin{figure}[!htbp]
\begin{center}
\includegraphics[height=2.5in]{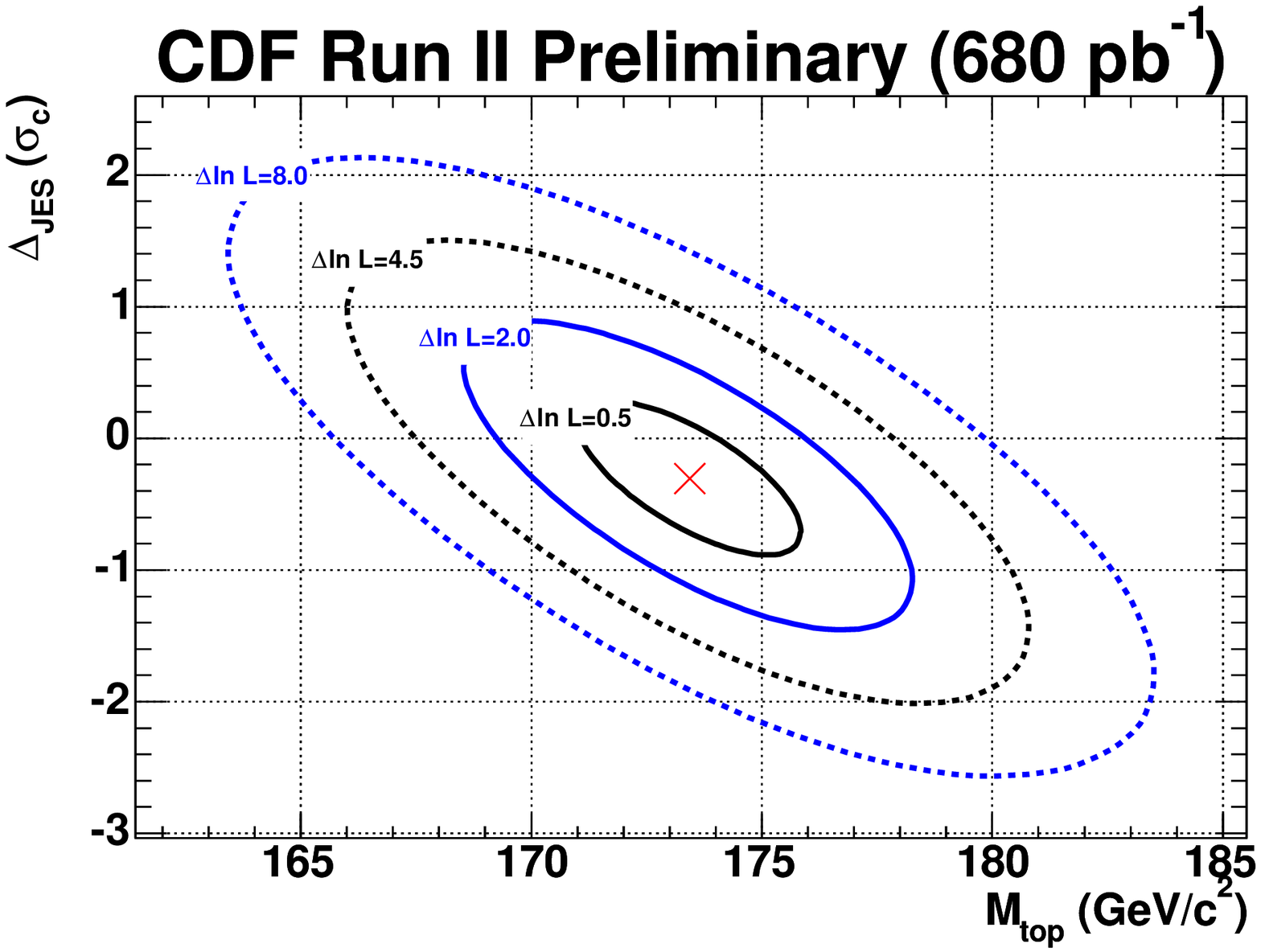}
\includegraphics[height=2.3in]{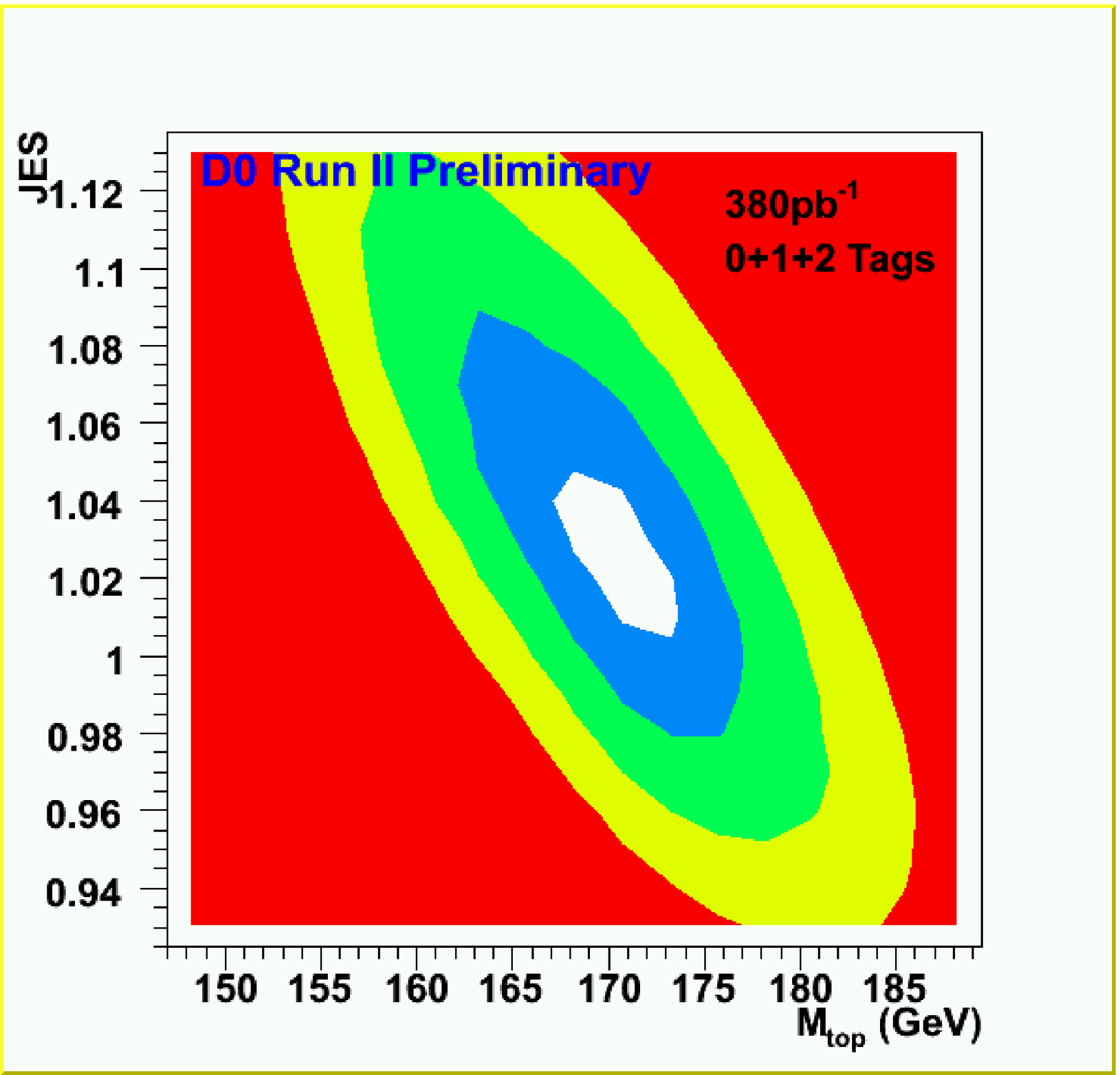}
\end{center}
\caption{ Left, the most precise single measurement of the top quark mass. Made in the lepton+jets channel at CDF~\cite{cdfljets}, it uses a template method and simultaneously fits the jet energy scale from $W\rightarrow qq$ decays. Right,  a measurement at D\O\ in the sample channel~\cite{d0ljets} using a matrix-element method and simultaneously fitting the jet energy scale from $W\rightarrow qq$ decays.}
\label{fig:ljets}
\end{figure}

The most precise single measurement of the top quark mass\cite{cdfljets} has been made in the lepton+jets channel using 680 pb$^{-1}$ of data at CDF (Figure~\ref{fig:ljets}) with a template method.  The templates are two-dimensional and parameterized in terms of the reconstructed top quark mass and reconstructed hadronic $W$ mass. This allows simultaneous fits to the top quark mass and the the jet energy scale (JES) from the $W\rightarrow qq'$ portion of the hadronic decay, yielding:

\[ M_{t} = 173.4\pm 1.7_{stat}\pm 1.8_{JES}\pm1.3_{syst} {\textrm GeV}/c^2  \]
\noindent

   A measurement from D\O\ in the same channel with 370 pb$^{-1}$~\cite{d0ljets} uses a matrix-element technique (Figure~\ref{fig:ljets}) and simultaneously fits the jet energy scale as well:

\[ M_{t} = 170.6 \pm 4.4_{stat} \pm 1.4_{syst} {\textrm GeV}/c^2,  \]
\noindent
where the statistical error includes the JES uncertainty. CDF has made two other measurements in the lepton+jets channels which give consistent results~\cite{cdfmeat,cdflxy}.

\section{Dilepton Measurements}

\begin{figure}[!htbp]
\begin{center}
\includegraphics[width=3in]{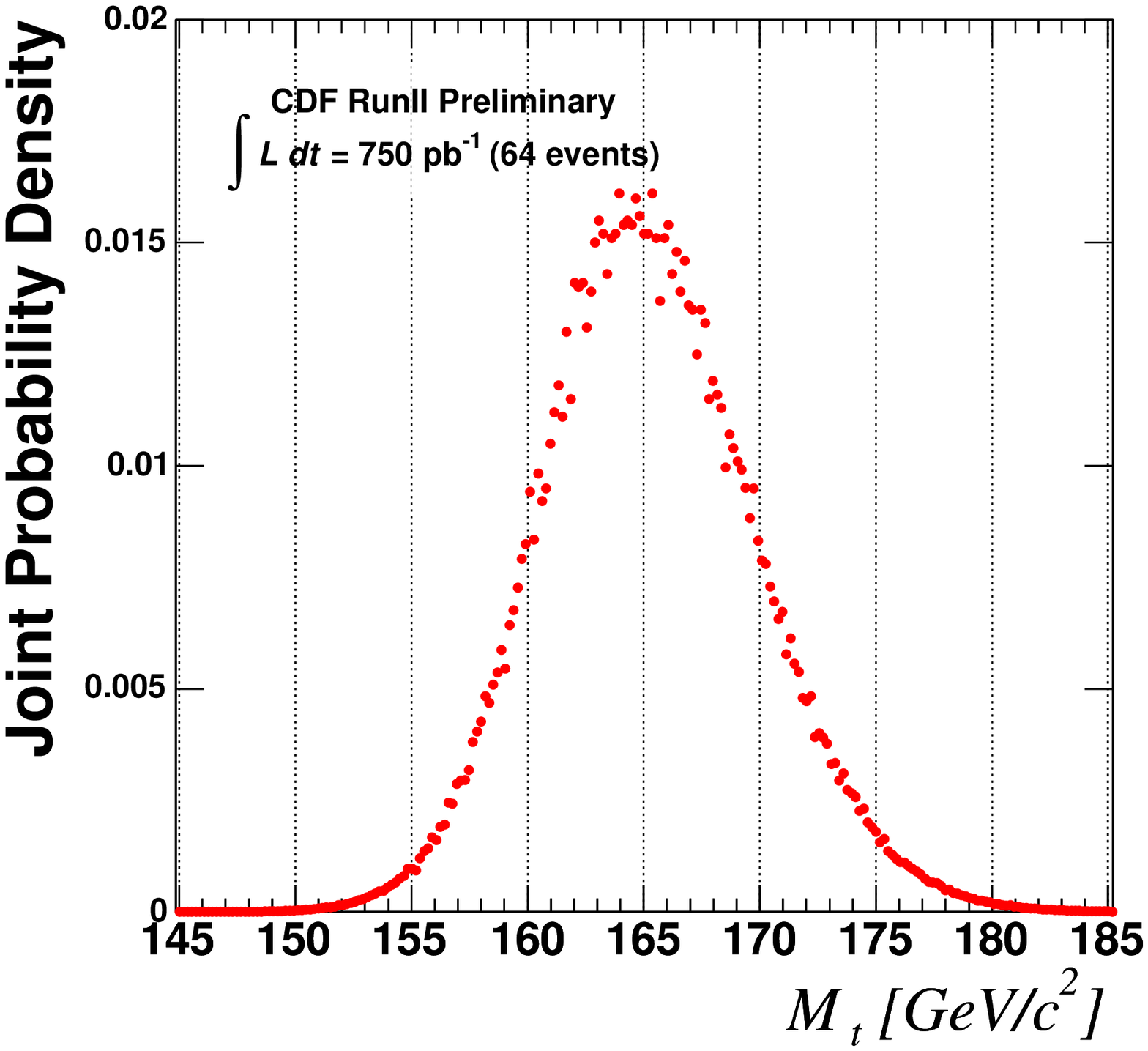}
\includegraphics[width=2.5in]{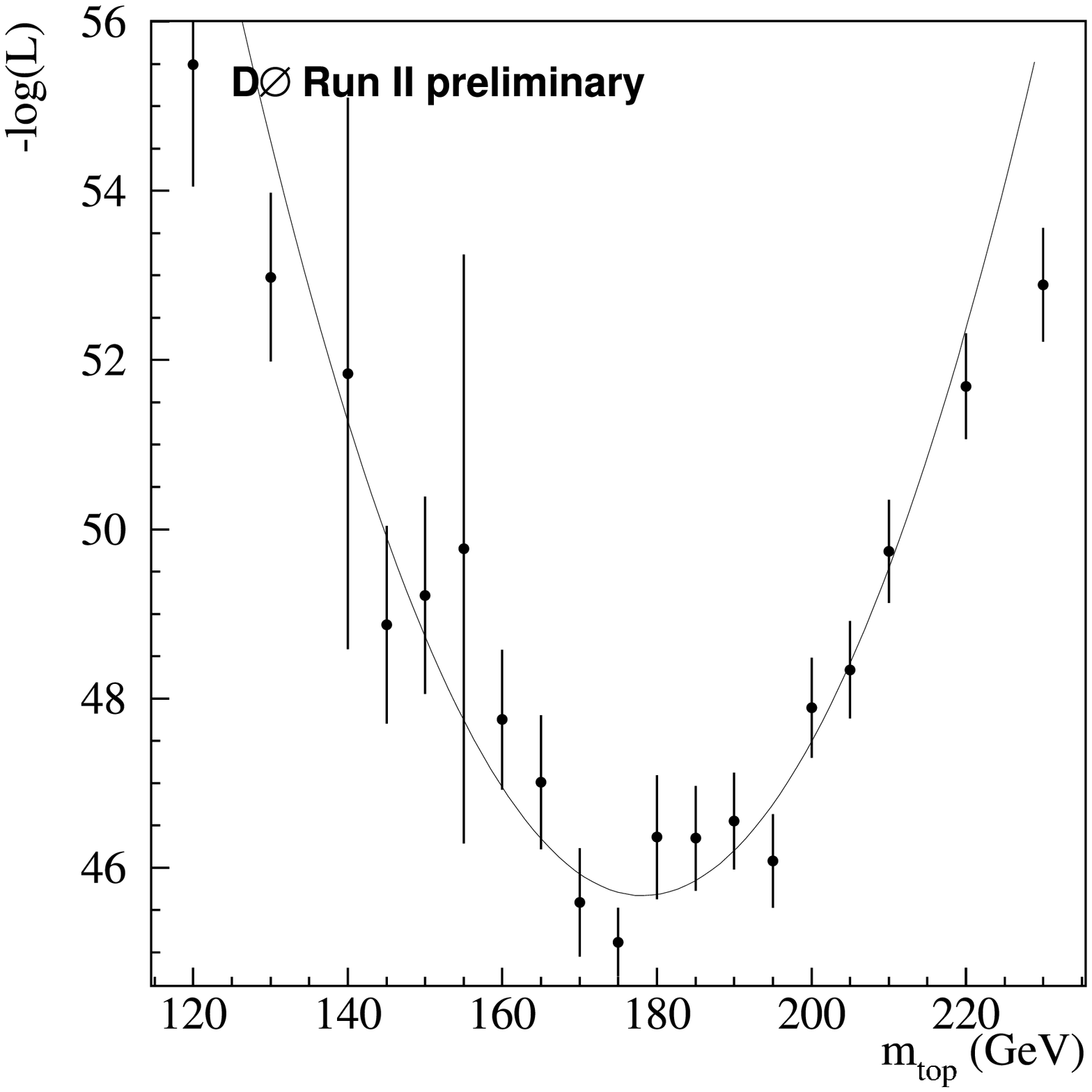}
\end{center}
\caption{ Left, a measurement of the top quark mass in the dilepton channel at CDF using a matrix-element method. Right, a measurement of the top quark mass in the same channel at D\O\ using a template method. }
\label{fig:dil}
\end{figure}

The dilepton channel has a smaller branching ratio, but is less reliant on the calibration of the jet energy scale.   An application of the matrix element method to 750 pb$^{-1}$ of data at CDF (Figure~\ref{fig:dil}) yields the most precise single measurement~\cite{cdfdil} in the dilepton channel:

\[ M_{t} = 164.5 \pm 4.5_{stat} \pm 3.1_{syst} \  {\textrm GeV}/c^2  \]

D\O\ has measured the top mass in the dilepton channel using a template method (Figure~\ref{fig:dil}) in a 370 pb$^{-1}$ sample which requires a tagged $b$-jet~\cite{d0dil}. The templates are parameterized in terms of a most likely reconstructed mass calculated by integrating the final state parton momenta and weighting by the experimental resolution. This measurement gives:

\[ M_{t} = 176.6 \pm 11_{stat} \pm 4_{syst }\  {\textrm GeV}/c^2  \]

D\O\ has made a second measurement in the dilepton channel which gives consistent results~\cite{d0dil2}.

\section{Conclusions}

\begin{figure}[!htbp]
\begin{center}
\includegraphics[width=3in]{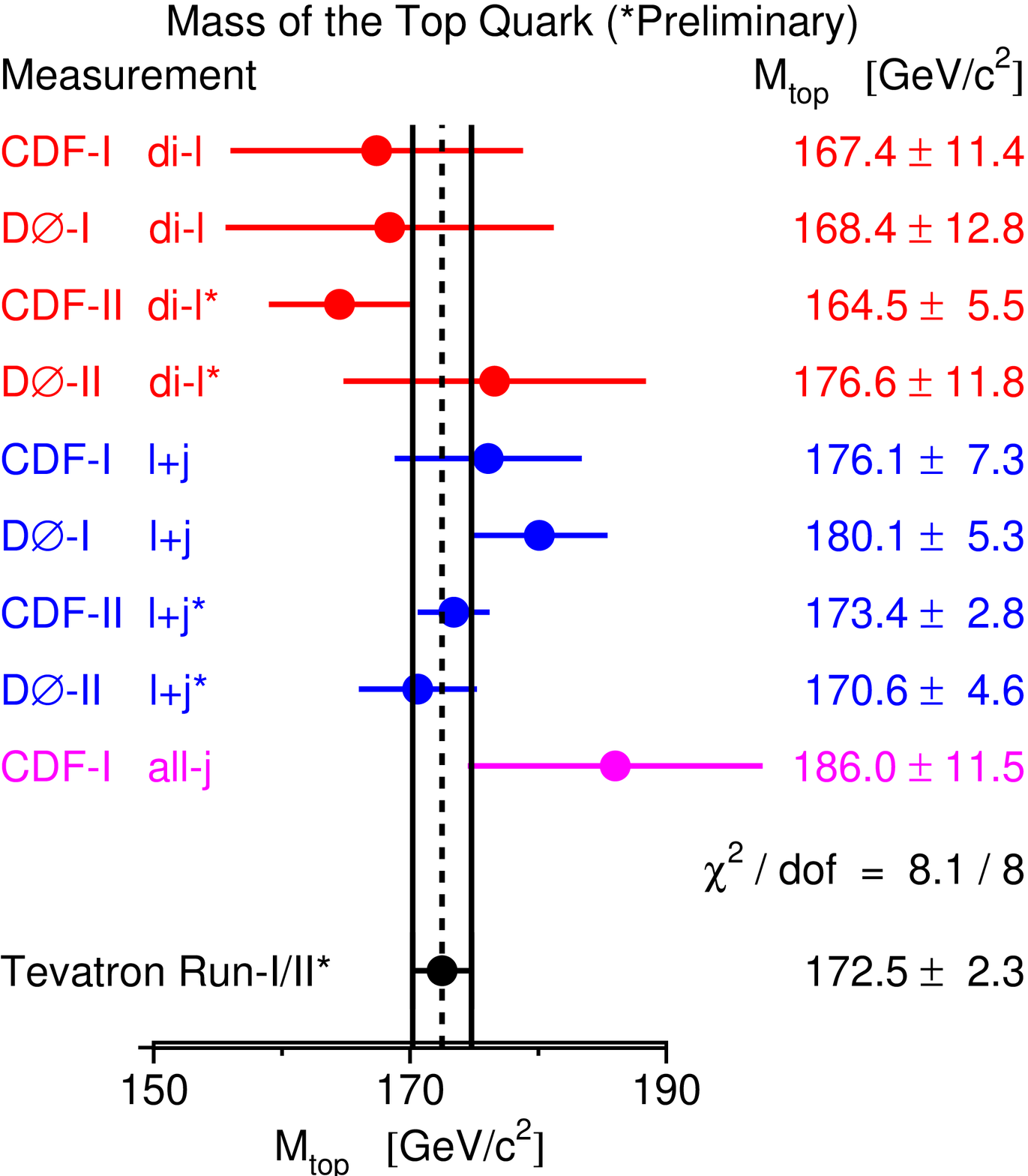}
\includegraphics[width=2.4in]{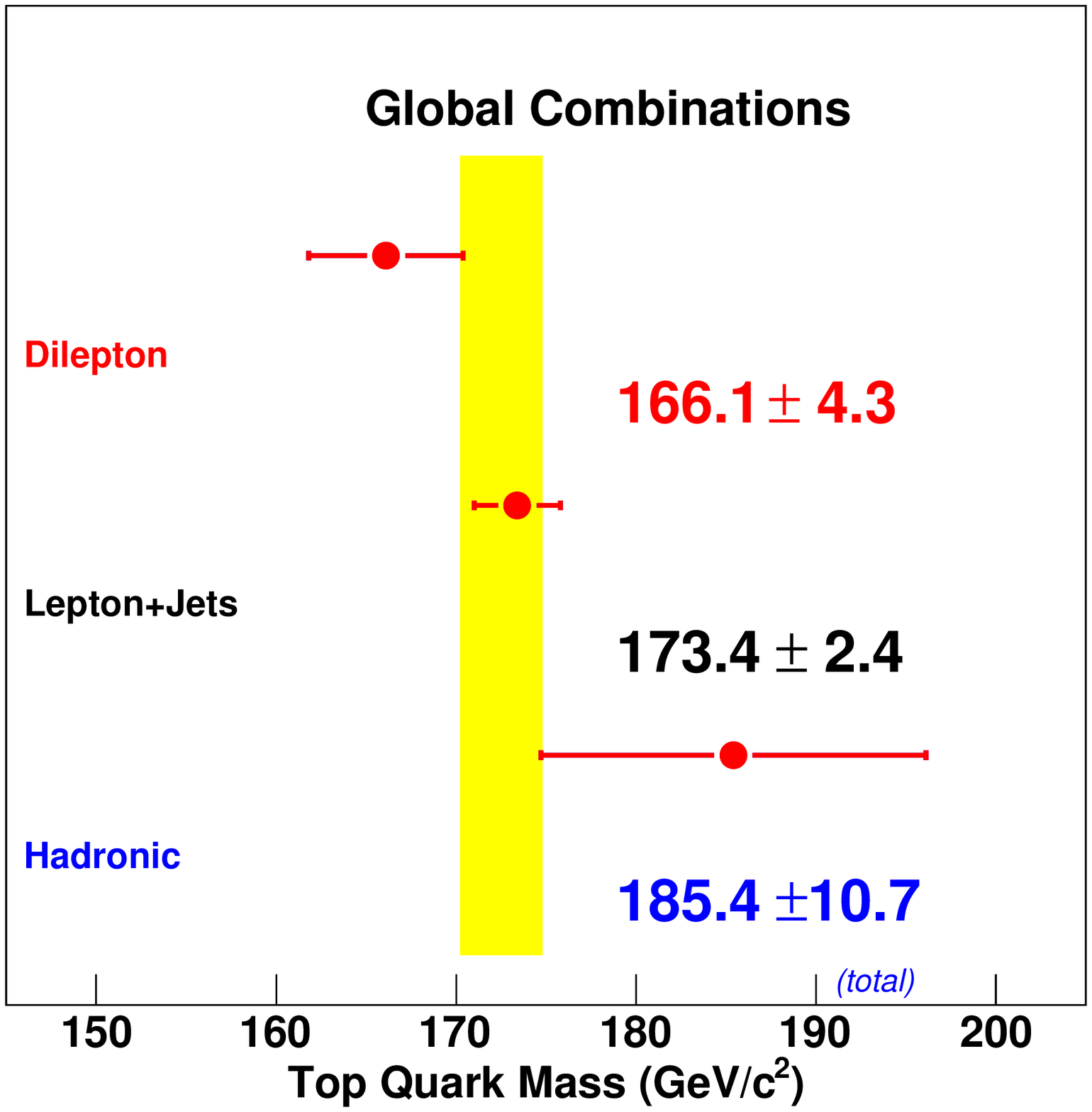}
\end{center}
\caption{ Left, combination of the most precise top quark mass measurements in each channel. Right, contributions to the global combination from each decay channel. }
\label{fig:combo}
\end{figure}

The most precise measurement in each channel from each experiment have been combined into a global average~\cite{combo} which yields the most precise determination of the top quark mass, see Figure~\ref{fig:combo}:

\[ M_{top} = 172.5 \pm 1.3_{stat} \pm 1.9_{syst} {\textrm GeV}/c^2.  \]

Direct measurements of the top quark mass at the Tevatron have yielded a precision of 2.3 GeV$/c^{2}$ in less than 1 fb$^{-1}$ of data. The measurements in the dilepton channel have tended historically to be slightly lower than those in the lepton+jets channel. The most recent precision measurement in the dilepton channel continues this trend, as seen in a comparison of the contributions of each channel to the global average (Figure~\ref{fig:combo}).  However, the combination of precision measurements has a $\chi^2/d.o.f = 8.1/8$, which is consistent with statistical fluctuations.  

\begin{figure}[!htbp]
\begin{center}
\includegraphics[width=4in]{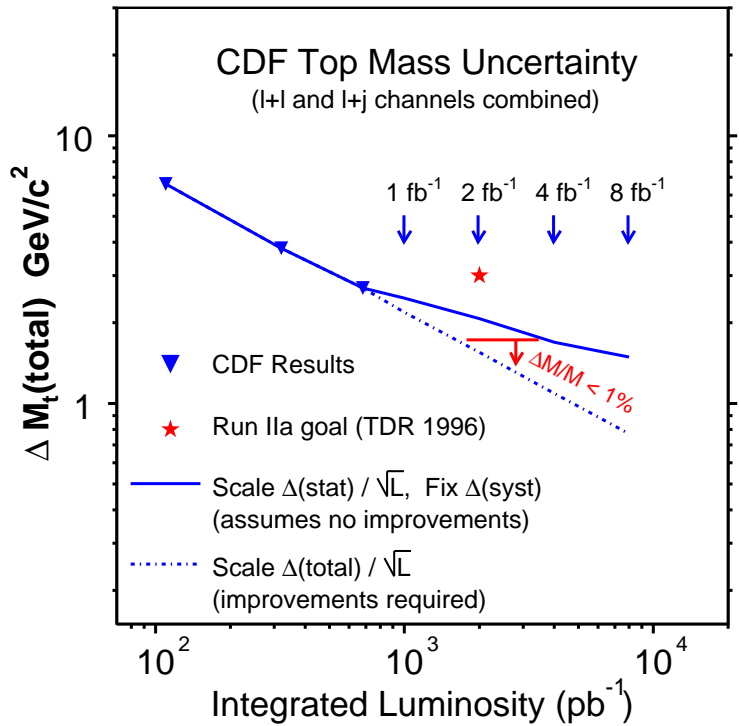}
\end{center}
\caption{ Extrapolation of top mass uncertainty at CDF to higher luminosity samples, with a scenario in which only the statistical error scales with luminosity, and a more optimistic scenario in which the systematic errors are reduced as well in the higher luminosity sample.}
\label{fig:extrap}
\end{figure}

In larger samples, CDF expects measurements in all three channels to become limited by systematic uncertainties, allowing a significant comparison of the measured mass in each channel and a total precision of less than 1\% (Figure~\ref{fig:extrap}).

\end{document}